\RequirePackage[2020-02-02]{latexrelease}
\documentclass[twocolumn,english,aps,prl,10pt,superscriptaddress]{revtex4}
\usepackage[T1]{fontenc}
\usepackage{amsmath,graphicx,amssymb,epsfig,babel,dsfont}
\usepackage{mathrsfs}
\usepackage{color}
\usepackage{amsbsy}
\usepackage{pdfpages}
\usepackage{pgffor}

\def\bbm[#1]{\mbox{\boldmath $#1$}}

\begin{document}

\title{Radiative heat exchange driven by acoustic vibration\\modes between two solids at the atomic scale}

\author{M. G\'omez~Viloria}
\affiliation{Laboratoire Charles Fabry, UMR 8501, Institut d'Optique, CNRS, Universit\'{e} Paris-Saclay, 2 Avenue Augustin Fresnel, 91127 Palaiseau Cedex, France.}

\author{Y. Guo}
\affiliation{Institut Lumière Matière, Universit\'e Claude Bernard Lyon 1, CNRS, Universit\'e de Lyon, 69622 Villeurbanne, France.}

\author{S. Merabia}
\affiliation{Institut Lumière Matière, Universit\'e Claude Bernard Lyon 1, CNRS, Universit\'e de Lyon, 69622 Villeurbanne, France.}

\author{R. Messina}
\affiliation{Laboratoire Charles Fabry, UMR 8501, Institut d'Optique, CNRS, Universit\'{e} Paris-Saclay, 2 Avenue Augustin Fresnel, 91127 Palaiseau Cedex, France.}

\author{P. Ben-Abdallah}
\email{pba@institutoptique.fr}
\affiliation{Laboratoire Charles Fabry, UMR 8501, Institut d'Optique, CNRS, Universit\'{e} Paris-Saclay, 2 Avenue Augustin Fresnel, 91127 Palaiseau Cedex, France.}

\date{\today}

\begin{abstract}
When two solids are separated by a vacuum gap of thickness smaller than the wavelength of acoustic phonons, the latter can tunnel across the gap thanks to van der Waals forces or electrostatic interactions. Here we show that these mechanical vibration modes can also contribute significantly, at the atomic scale, to the nonlocal radiative response of polar materials. By combining molecular-dynamics simulations with fluctuational-electrodynamics theory we investigate the near-field radiative heat transfer  between two slabs due to this opto-mechanical coupling and we highlight its dominant role at cryogenic temperatures. These results pave the way to exciting avenues for the control of heat flux and the development of cooling strategies at the atomic scale.
\end{abstract}

\maketitle

The physics of heat transfer between two solids separated by a vacuum gap in the transition regime between conduction and radiation remains today largely unknown. When solids are separated by gaps having thicknesses of tens of nanometres or more heat transfer is exclusively driven by photons exchange. In the far-field regime (distances larger than the thermal wavelength, around 10$\,\mu$m at ambient temperature) this transfer is limited by Stefan-Boltzmann's law defining the blackbody limit~\cite{Planck}. At subwavelength scale and down to distances of about ten nanometers, the heat flux exchanged between the solids can overcome this limit by several orders of magnitude~\cite{Rytov,Polder,Joulain,Volokitin,RMP, Hargreaves,Narayanaswamy,Shen,Rousseau,Ottens,Kralik} thanks to the tunneling of evanescent photons which superimposes to the flux driven by propagative photons. Below this separation distance, heat transfer can be mediated by multiple carriers~\cite{arxiv,Francoeur1,Francoeur2,Guo,MGV}. More specifically, at sub-nanometer scale, acoustic vibration modes of solids participate in the transfer. In 2015, Chiloyan et al. highlighted~\cite{Chiloyan}, by means of atomistic simulations, the dominant role played by these modes on the transfer between polar materials. In that work, it was claimed that this transfer results from the tunneling of vibration modes thanks to surface forces. Such a transfer has been described theoretically in the continuum limit for isotropic media by Pendry et al.~\cite{Pendry} (see also Refs.~\cite{Volokitin2,Volokitin3}) using the classical elasticity theory. More recently this description has been extended to anisotropic piezoelectric materials~\cite{Maasilta}.
In the present Letter we show that acoustic vibration modes, which are traditionally purely mechanical modes in the long-wavelength (LW) limit, are not only able to tunnel through the separation gap thanks to the surface forces existing between the two solids, but also contribute, at atomic-scale  separation distances (i.e. short wavelengths limit), to the nonlocal radiative response of materials. In the LW limit, it is well-known that the optical phonons are the only excitations that give rise to local electric dipole moments inside the material owing to the motion of neighboring atoms with opposite partial charges in opposite directions. In this limit, the optical phonons are the only link between the atomic vibrations within the solid and the surrounding electromagnetic field. These optical vibration modes entirely drive the radiative response of material. Although a mesoscopic theory describing the nonlocal response of polar materials has been recently introduced~\cite{De Liberato} in an analogous way as the hydrodynamic description of electron gas in nonlocal plasmonics~\cite{Fuchs,Cirac}, this theory suffers from a fundamental limitation to properly describe light-matter interactions at the atomic scale. Indeed, it ignores the crucial role played by the acoustic vibration modes in the radiative response of material.
However, as we will see, at atomic scale, acoustic vibration modes play a major role in this response. Here, we make a detailed description of this opto-mechanical coupling  and we highlight its importance in the radiative heat exchanges between two solids close to the physical contact. We demonstrate that, contrary to wide belief, these acoustic vibration modes can significantly contribute to the radiative heat exchanges and can even be dominant, in front  of the contribution coming from optical phonons, in the cryogenic regime.

\begin{figure*}[t!]
	\includegraphics[width=0.26\linewidth]{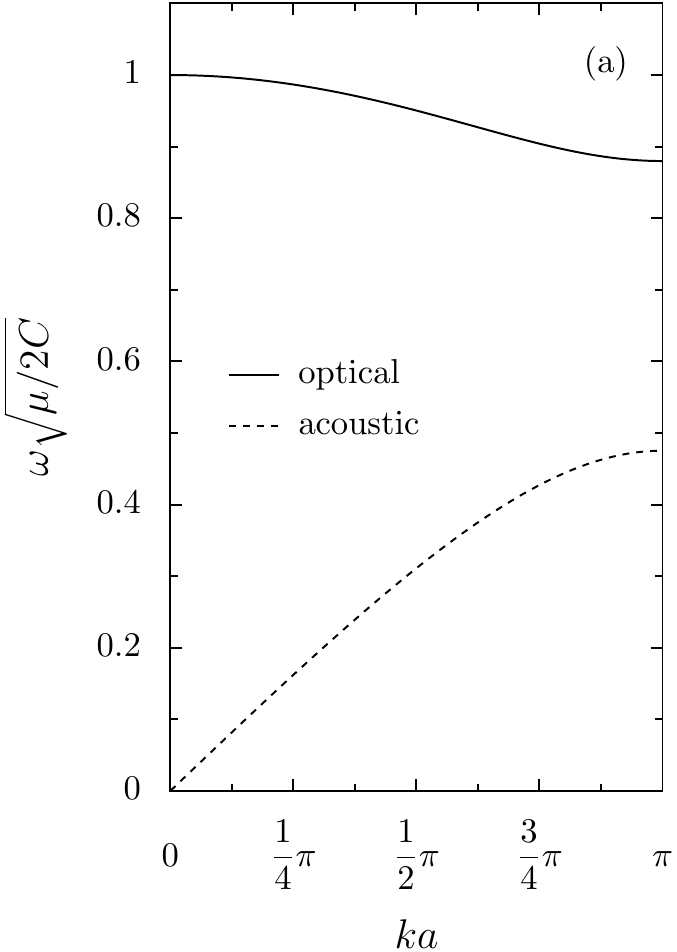}\hspace{.5cm}\includegraphics[width=0.36\linewidth]{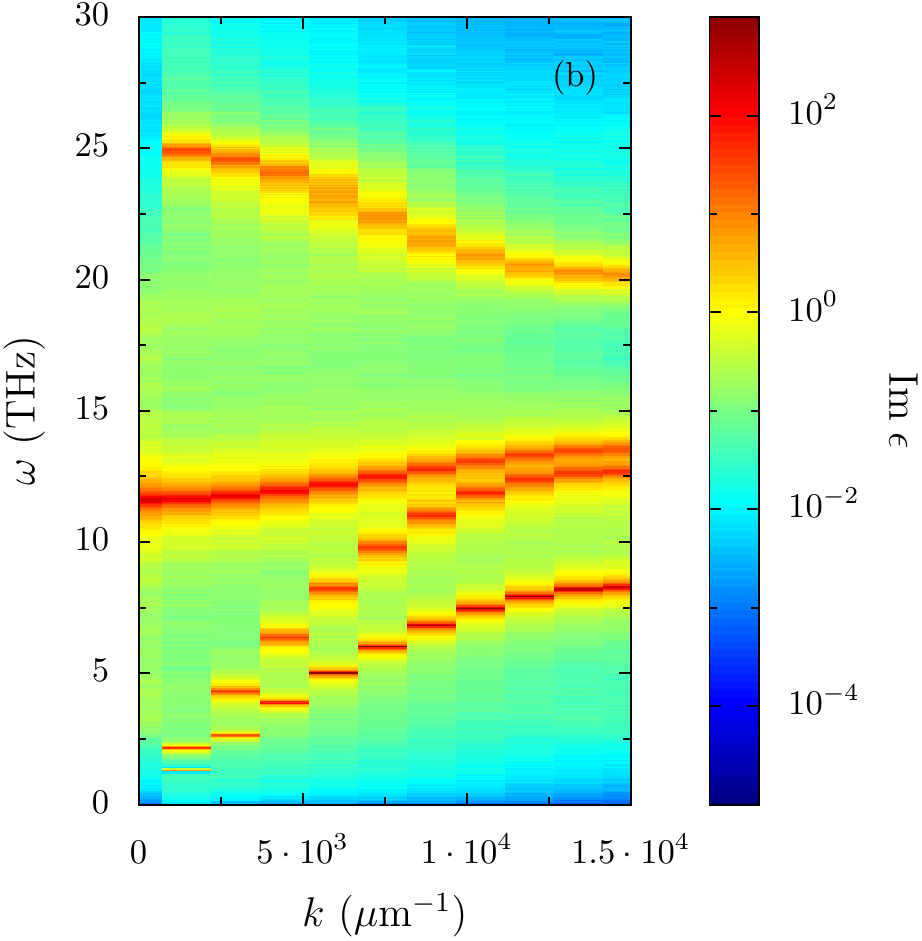}\hspace{.5cm}\includegraphics[width=0.27\linewidth]{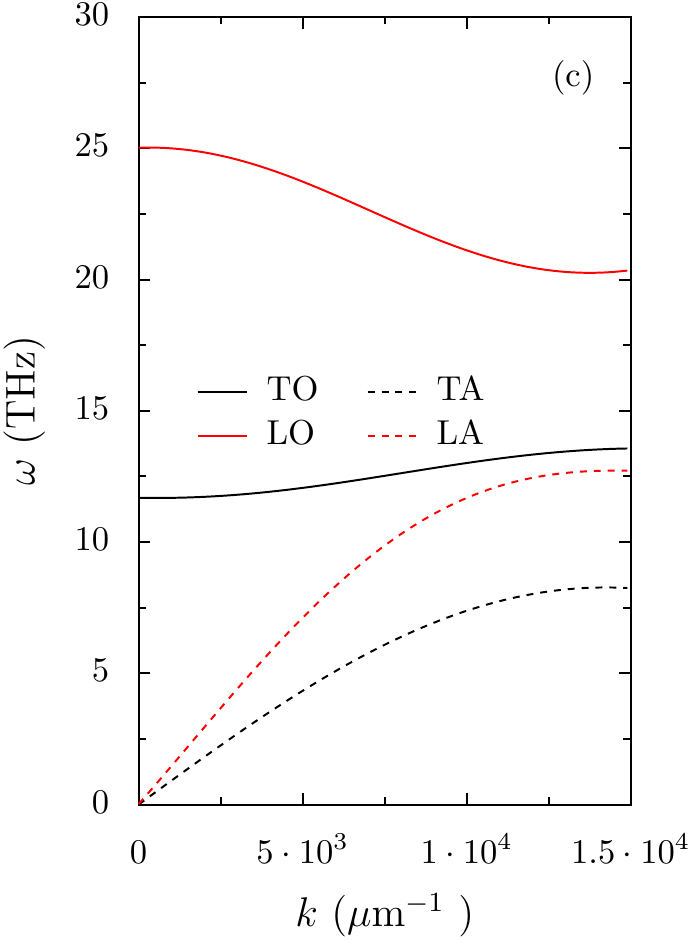}
	\caption{(a) Dispersion relation of eigenmodes in a diatomic chain with a reduced mass $\mu=\frac{M_1M_2}{M_1+M_2}$, chain stiffness $C$ and lattice constant $a$. (b) Imaginary part of the dielectric permittivity of a MgO cubic crystal (bulk) in the (001) direction of reciprocal space, calculated by molecular dynamics. (c) Dispersion relation of eigenmodes in the MgO crystal obtained by solving its secular equation. The four eigenmodes are known as the longitudinal optical (LO; solid red), transverse optical (TO; solid black), longitudinal acoustic (LA; dashed red) and transverse acoustic (TA; dashed black) branches.}
	\label{Fig1}
\end{figure*}

To start, let us consider a polar material of arbitrary crystalline structure. To describe light interaction with this crystal in the infrared frequency range and at atomic-scale separation distances, a nonlocal model of the radiative response of the material must be introduced.  To go beyond the Gubbin and De Liberato theory~\cite{De Liberato} a fully atomistic calculation of the dielectric response has been performed (see Supplemental Material \cite{suppmat} and references ~\cite{thompson,matsui,chalopin,Callen} therein), based on the analysis of the fluctuations of the polarization density within the crystal. In spatially invariant crystals this density is related to local electric field $\bold{E}(\bold{r},t)$ through the nonlocal relation (assuming the system stationary)
\begin{equation}
	\bold{P}(\bold{r},t)=\epsilon_0\int\int \mathrm d\bold{r'}\mathrm dt'\;\stackrel{\leftrightarrow}{\chi}\!(\bold{r}-\bold{r'},t-t')\cdot\bold{E}(\bold{r'},t'),
	\label{linear relation}
\end{equation}
where $\epsilon_0$ is the vacuum permittivity, whereas $\stackrel{\leftrightarrow}{\chi}\!(\bold{r}-\bold{r}',t-t')$ denotes the electric susceptibility tensor of the crystal,  $\bold{r}-\bold{r}'$ and $t-t'$ being the spatial and temporal distance, respectively, between polarization and electric field. According to the fluctuation-dissipation theorem~\cite{Chandler}, the spatial Fourier components of the susceptibility at temperature $T$ read
\begin{equation}
	\chi_{mn}(\bold{k},t)=-H(t)\frac{1}{\epsilon_0 k_{\rm B} T}\frac{\mathrm d}{\mathrm dt}\langle \delta P_m(\bold{k},0)\delta P^*_n(\bold{k},t)\rangle,
	\label{FDT}
\end{equation}
where $\delta P_m=P_m-\langle P_m \rangle$, $\langle .\rangle$ denotes an ensemble average,  $*$ is the conjugation operation and $H$ is the Heaviside function. For a statistically stationary process (i.e. $\mathrm d\langle P_m \rangle/\mathrm dt=0$) we have equivalently 
\begin{equation}
	\chi_{mn}(\bold{k},t)=-H(t)\frac{1}{\epsilon_0 k_{\rm B} T}\frac{\mathrm d}{\mathrm dt}\langle P_m(\bold{k},0) P^*_n(\bold{k},t)\rangle.
	\label{FDT2}
\end{equation}
It turns out by time Fourier transformation of this expression that the electric susceptibility and the relative dielectric permittivity of the crystal read
\begin{equation}
\begin{split}
	\chi_{mn}(\bold{k},\omega)&=\frac{1}{\epsilon_0 k_{\rm B} T}\bigg[\langle P_m(\bold{k},0)P^*_n(\bold{k},0)\rangle\\
 &\,\left.+\,\mathrm i\omega\int_0^{\infty}\mathrm dt\; e^{\mathrm i\omega t}\langle P_m(\bold{k},0)P^*_n(\bold{k},t)\rangle \right]
\end{split}
	\label{susceptibility2}
\end{equation}
and $\epsilon_{mn}(\mathbf k, \omega)=\chi_{mn}(\mathbf k, \omega)+\delta_{mn}$, respectively, $\delta_{mn}$ being the Kronecker delta. These expressions relate the nonlocal radiative response of material to the correlations functions of local dipolar moments. The latter are in turn related to the motion of partial charges which are associated to each atom. In contrast, we define the local response as the limit $\chi_{mn}(\mathbf k,\omega)\to \chi_{mn}(\mathbf k=0,\omega)$ valid for large separation distances. Relation (\ref{susceptibility2}) can be generalized to the quantum regime by relating the correlation function of fluctuating polarization density operator to the electrical susceptibility through the general Kubo formula~\cite{Zwanzig}.

To give insight into the link between the vibration modes and the radiative response of material we detail below the case of the diatomic chain~\cite{Kittel} which is the simplest polar crystal. In this particular case, it is well known that the dispersion relation of resonant modes reads
\begin{equation}
\begin{split}
	\omega^2 &=C\left(\frac{1}{M_1}+\frac{1}{M_2}\right)\\
	&\,\pm C\left[\left(\frac{1}{M_1}+\frac{1}{M_2}\right)^2-\frac{4\sin^2(ka/2)}{M_1 M_2}\right]^{1/2},
	\label{dispersion}
\end{split}
\end{equation}
where $C$ denotes the chain stiffness between the atoms of mass $M_1$ and $M_2$ while $a$ is the lattice period and $k$ is the mode wavenumber. As for the amplitudes $u_l=u_k \exp(\mathrm{i}[kla-\omega t])$ and $v_l=v_k \exp(\mathrm{i}[kla-\omega t])$ of the normal modes associated with the masses $M_1$ and $M_2$, respectively, in the unit cell $l$ they satisfy the relation
\begin{equation}
	\frac{u_k}{v_k}=\frac{2Ce^{-\mathrm ika/2}\cos (ka/2)}{2C-M_1\omega^2}.
	\label{amplitude}
\end{equation}
In the LW limit ($k\to0$), the dispersion relation of optical (high frequency) and acoustic (low frequency) branches read
\begin{equation}
	\omega^2=2C\left(\frac{1}{M_1}+\frac{1}{M_2}\right), \qquad \omega^2=\frac{C}{2(M_1+M_2)}(k a)^2, 
	\label{dispersion2}
\end{equation}
and the amplitudes of optical and acoustic normal modes satisfy respectively the relations $u_k/v_k\approx -M_2/M_1$ (i.e. out-of-phase atomic vibration) and $u_k/v_k\approx 1$ (i.e. in-phase atomic vibration), showing that only the optical modes give rise to dipole moments.
On the other hand, close to the upper bound of the Brillouin zone (i.e. $k\approx\pi/a$), that is in the extreme near-field regime, the situation radically changes. As shown in Fig.~\ref{Fig1}(a), an anticrossing of acoustic and optical branches appears in this zone showing a strong coupling between these modes with a frequency splitting (assuming here $M_2<M_1$) $\Gamma=\omega_{\rm o}-\omega_{\rm a}$, with $\omega_{\rm o}=\sqrt{2C/M_2}$ and $\omega_{\rm a}=\sqrt{2C/M_1}$.
As far as the amplitudes of normal modes are concerned, we see from the general expression \eqref{amplitude} that 
\begin{equation}\frac{u_k}{v_k}\approx\frac{\mathrm iC(ka-\pi)}{2C-M_1\omega^2_{\rm o,a}},
\end{equation}
so that $u_k/v_k\to0$ for the optical modes and $u_k/v_k\to\infty$ for the acoustic modes. These relations demonstrate that both types of modes give rise to dipole moments and therefore both contribute to the radiative response of the chain. Moreover these relations also demonstrate that for both acoustic and optical modes one atom is motionless in the unit cell while the second is free to oscillate making these modes identical in nature. 

\begin{figure*}[t!]
	\includegraphics[width=0.4\linewidth]{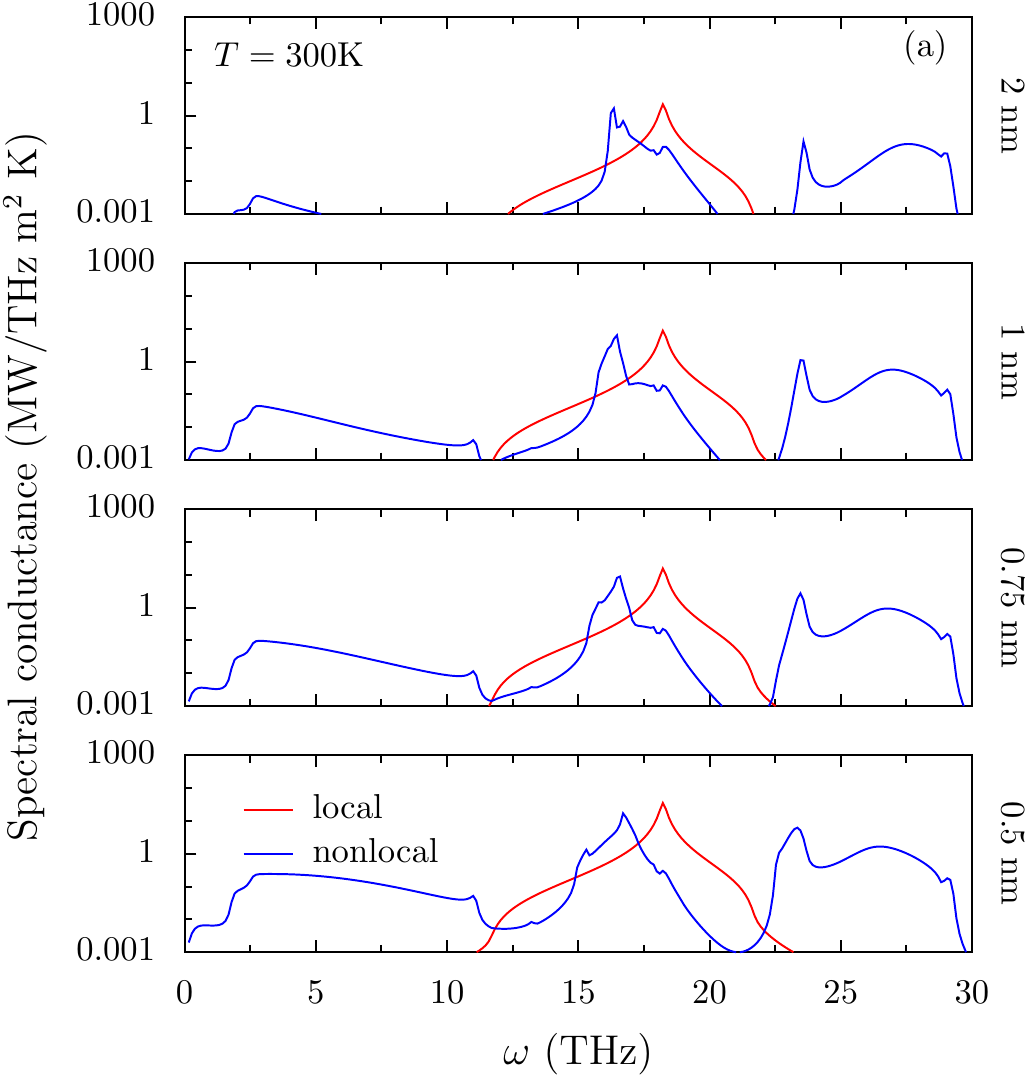}\hspace{1cm}
	\includegraphics[width=0.4\linewidth]{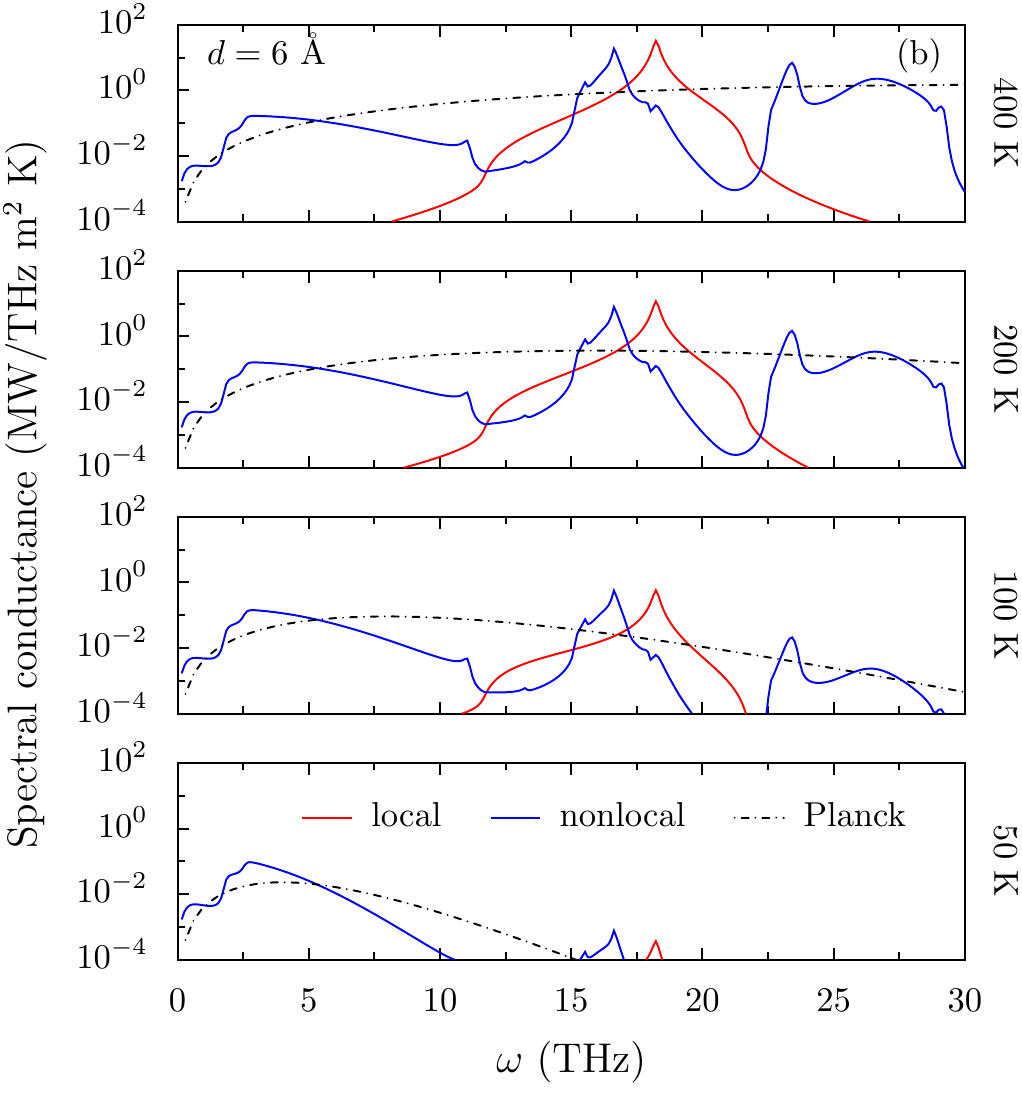}
	\caption{Thermal conductance spectra between two MgO samples for different separation distances at (a) $T=300$~K and (b) for different temperatures at $d=6$~\AA. The conductance is calculated using both the local (red) and nonlocal (red) dielectric response of material.  The dashed curves represent Planck's law in arbitrary units.}
	\label{Fig2}
\end{figure*}

The radiative contribution of acoustic modes to the nonlocal radiative response can be directly observed in a concrete scenario. In Fig.~\ref{Fig1}(b) we show the dielectric response of a magnesium oxide (MgO) crystal in the $(001)$ axis of reciprocal space obtained by molecular-dynamics simulation~\cite{suppmat}. This material has been chosen to get well-separated optical and acoustic branches making the analysis and interpretation of results easier. The comparison of this mapping with the dispersion relations of resonant vibration modes [Fig.~\ref{Fig1}(c)] calculated by solving the secular equation of the crystal clearly shows the contribution of optical branches at high frequencies but also the one of acoustic branches at low frequencies. We also observe, at the edge of the Brillouin zone, the anticrossing, previously mentioned for the diatomic chain, between the longitudinal optical (LO) and the longitudinal acoustic (LA) branches, true signature of strong coupling between these vibration modes in this region. It is worthwhile to note that the contribution of acoustic modes is not limited to the edge of the Brillouin zone. In Fig.~\ref{Fig1}(b) we see that the acoustic mode can couple to the electromagnetic field relatively far away from this region. For MgO, the contribution of acoustic modes to the nonlocal response of crystal can be observed down to wavectors $k\approx 1/2a$, $a=4.2$~\AA{} being the lattice constant of crystal. 

The role played by these modes on the radiative heat transfer can then be investigated thanks to fluctational-electrodynamics theory. According to this framework, the conductance of radiative heat exchanged at temperature $T$ between two solids separated by a vacuum gap of thickness $d$ can be written in the Landauer-like form~\cite{Polder,pba,Biehs}
\begin{equation}
G(T,d)=\int_0^{\infty} \!\frac{\mathrm{d}\omega}{2\pi} \frac{\mathrm d\Theta}{\mathrm dT} (\omega,T)\int\!\frac{\mathrm{d} \boldsymbol{\kappa}}{(2\pi)^2}\; \sum_{\alpha=\mathrm{s},\mathrm{p}}\mathcal{T}_\alpha (\bold{\kappa},\omega, d),
\label{conductance}
\end{equation}
where $\Theta(\omega,T)={\hbar\omega}/[\exp(\hbar\omega/k_{\rm B} T)-1]$ is the mean energy of Planck oscillator at temperature $T$ and $\mathcal{T}_{\alpha}(\bold{\kappa},\omega,d)$ is the transmission coefficient in polarization $\alpha\in\{\mathrm{s,p}\}$ of mode ($\bold{\kappa},\omega$), $\bold{\kappa}$ being the parallel component of the wavector. Assuming a system with azimuthal symmetry this coefficient reads 
\begin{equation}\begin{split}
	\label{eq:transmission_rad}
	&\mathcal{T}_{\alpha}(\kappa,\omega,d)\\
	&\,=\begin{cases} 
		\displaystyle\frac{(1-|r_{\alpha,1}|^2)(1-|r_{\alpha,2}|^2)}{|1-r_{\alpha,1}r_{\alpha,2}\exp[2\mathrm i k_z d ]|^2}, & \kappa<\omega/c,\\\vspace{-0.3cm}\\
		\displaystyle\frac{4\,\mathrm{Im}\,r_{\alpha,1}\mathrm{Im}\,r_{\alpha,2}\exp[-2\,\mathrm{Im}\,(k_z) d]}{|1-r_{\alpha,1}r_{\alpha,2}\exp[-2\,\mathrm{Im}\,(k_z) d ]|^2}, & \kappa\geq \omega/c. \\
	\end{cases}
\end{split}\end{equation}
Here, $r_{\alpha,i}$ denotes the reflection coefficient of medium $i=1,2$ from vacuum and $k_z=\sqrt{(\omega/c)^2-\kappa^2}$ is the normal component of wavector in vacuum while $\kappa=|\boldsymbol{\kappa}|$. 
The reflection coefficients can be written in terms of surface impedances $Z_{\alpha, i}$ as follows~\cite{Ford}
\begin{subequations} 
\begin{equation}\label{eq:r_s}
		r_{\mathrm s, i} (\kappa,\omega)=\frac{\displaystyle Z_{\mathrm s, i}(\kappa,\omega)-\frac{\omega}{c^2k_z}}{\displaystyle Z_{\mathrm s, i}(\kappa,\omega)+\frac{\omega}{c^2k_z}},
\end{equation}
\begin{equation}
\label{eq:r_p}
	r_{\mathrm p, i}(\kappa,\omega)=\frac{\displaystyle \frac{k_z}{\omega}-Z_{\mathrm p, i}(\kappa,\omega)}{\displaystyle\frac{k_z}{\omega}+Z_{\mathrm p, i}(\kappa,\omega)},
\end{equation}
\end{subequations}
with~\cite{Esquivel}
\begin{subequations}
\begin{align}
	Z_{\mathrm s, i} (\kappa,\omega)=\frac{2\mathrm i}{\pi \omega}\int_0^\infty \frac{\mathrm{d}q_z}{\epsilon_{\mathrm t, i}(k, \omega)-(ck/\omega)^2},
\end{align}
\begin{align}
		&\nonumber Z_{\mathrm p, i} (\kappa,\omega)\\
		&\,=\frac{2\mathrm i}{\pi \omega}\int_0^\infty\frac{\mathrm{d}q_z}{k^2}\left[\frac{q_z^2}{\epsilon_{\mathrm t, i}(k, \omega)-(ck/\omega)^2}+\frac{\kappa^2}{\epsilon_{\mathrm l, i}(k, \omega)}\right],
\end{align}
\end{subequations}
where $k^2=q_z^2+\kappa^2$. Here $\epsilon_{\mathrm l, i}(k,\omega)$ and $\epsilon_{\mathrm t, i}(k,\omega)$, denote the longitudinal and transverse dielectric functions which are calculated by molecular-dynamics simulations~\cite{suppmat}. The computed spectra of heat conductances calculated from the integrand over the $\omega$-integral of expression (\ref{conductance}) with the nonlocal response of material   (dependent on wavevector {$k$}) for different separation distances and different temperatures are presented in (Figs.~\ref{Fig2}\:a,b) and compared with conductances calculated with the local dielectric permittivity $\epsilon(\omega)$  (independent of $k$).

\begin{figure}
\includegraphics[width=0.8\linewidth]{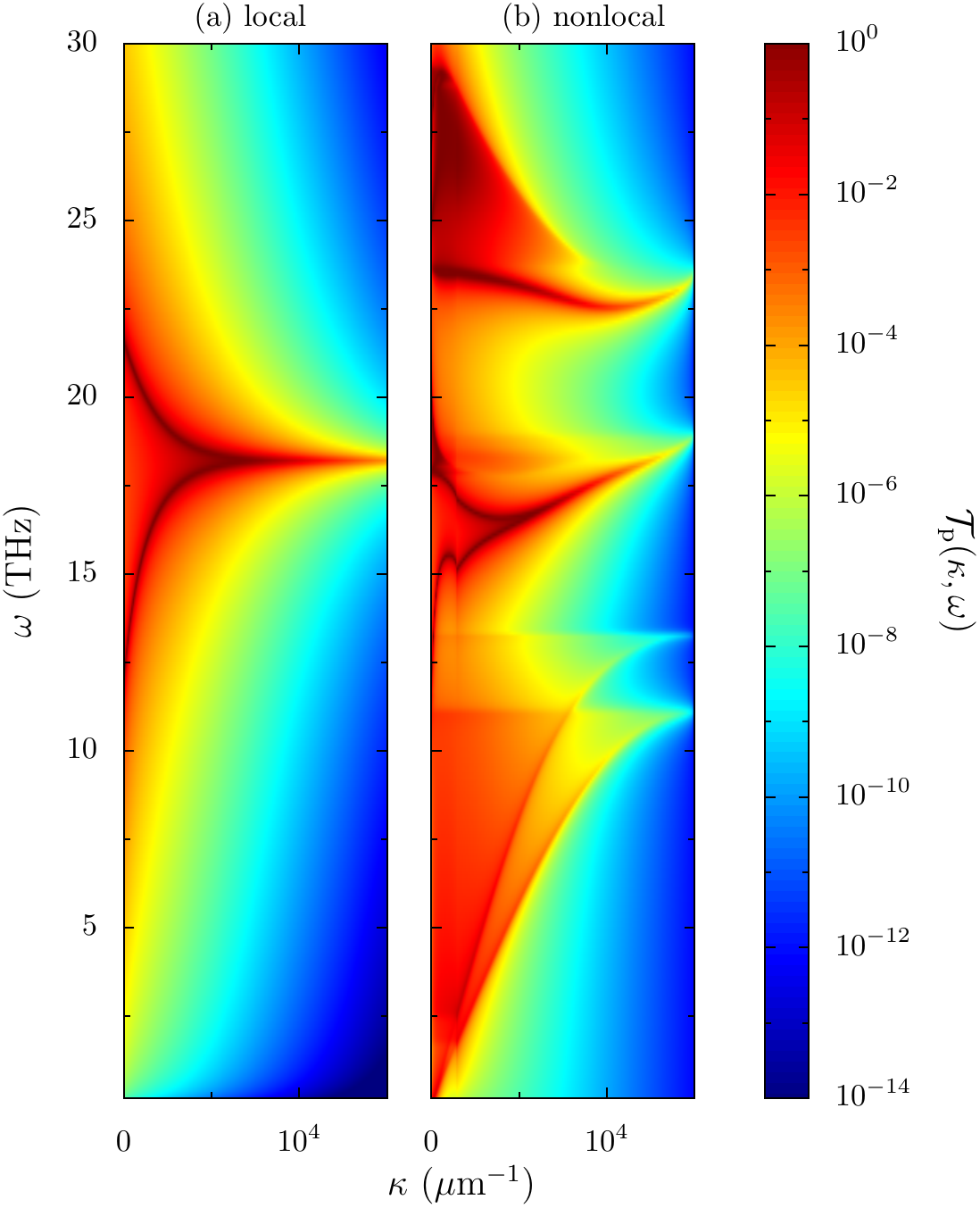}
\caption{Local (left) and nonlocal (right) transmission coefficient of p-polarized radiative heat exchange between two MgO samples separated by a vacuum gap of $d=6$~\AA{}}.
\label{Fig3}
\end{figure}

At gaps $d>1$\,nm we see that the heat transfer mainly stems from modes at high frequencies. The comparison of spectra with the dielectric permittivity plotted in Fig.~\ref{Fig1} shows that these modes are in the spectral range of optical phonons. However, below this critical distance we observe that the lower-frequency modes also participate in the transfer. We also note in Fig.~\ref{Fig2}(b) that the relative weight of these modes in comparison with the high-frequency modes increases at low temperature. These modes even become dominant in the cryogenic regime ($T<100$~K). The inspection of the transmission coefficients plotted in Fig.~\ref{Fig3} clearly shows that these low-frequency modes correspond to acoustic vibration modes.
 This result unquestionably demonstrates that the acoustic modes contribute radiatively to the transfer at small separation distances. Unlike the conductive-like heat transfer due to the mechanical tunneling of acoustic modes mediated by van der Waals forces between the two solids, this transfer is purely radiative and is related to dipole-dipole interactions induced by the acoustic vibration modes. Also, it must be noted (see Fig.\ref{Fig1}(b)) that the acoustic modes with very small wavevector do not play any role in the radiative response of the material. These modes can in principle participate to the heat transfer by tunneling but not to the radiative one. However, as shown in \cite{Chiloyan} this tunneling is negligible because of the weakness of surface forces.

\begin{figure}[t]
\includegraphics[width=0.9\linewidth]{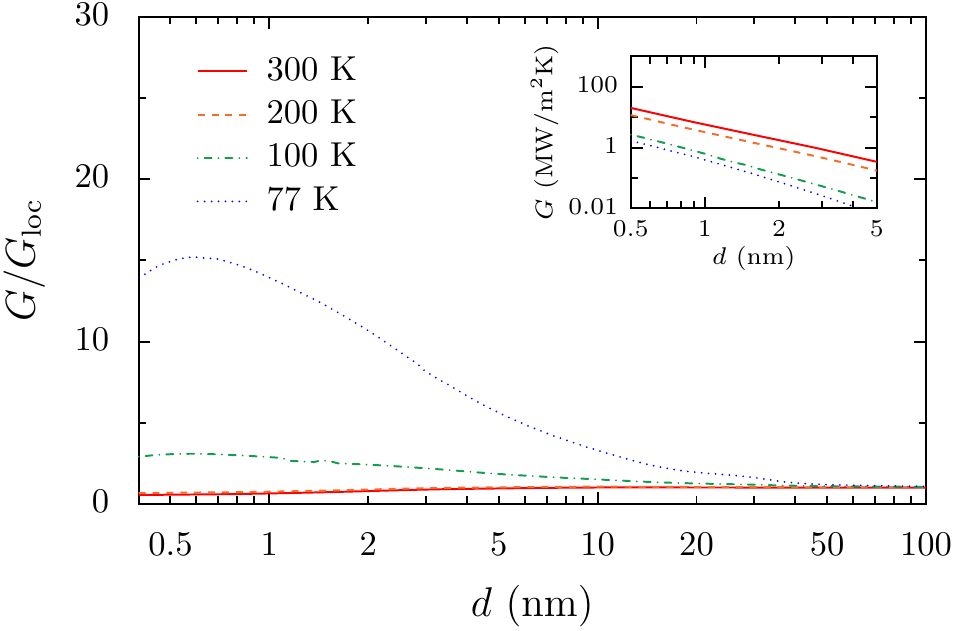}
\caption{Nonlocal vs local radiative conductance with respect to the separation distance for different temperatures. Inset: Full radiative conductance with respect to the separation distance.}
\label{fig:4}
\end{figure}

Finally, we analyze the ratio of the nonlocal radiative conductance to the local one.
The results plotted in Fig.~\ref{fig:4} with respect to the separation distance for different temperatures show, at large distance, that the local and non local conductances become, as expected, identical. Moreover we note that nonlocal effects dominate at low temperature, in the cryogenic regime owing to the dipoles generated by the acoustic vibration modes.
This result would be trivial if these modes were known to give rise to electric dipoles. But this is not generally the case for small wavevectors. Hence, at relatively large distances [see Fig. 2(a)] these modes do not contribute to the transfer. On the other hand, at close separation distance (large wavectors), this is not true anymore and these mechanical vibration modes contribute significantly to the heat transfer, this contribution being purely radiative in nature. In the inset of Fig.~\ref{fig:4} we see at ambient temperature that the conductance follows the usual power law scaling in $1/d^2$, $d$ being the separation gap between the two plates (see ~\cite{suppmat}). On the the other hand, in the cryogenic regime the scaling changes since the transfer is not mediated anymore by the surface phonon-polaritons.

In this work we shed light on the radiative heat transfer between polar materials close to the physical contact. We have shown that the acoustic vibration modes play a major role in the nonlocal radiative response of material and even become the dominant channel for radiative heat exchanges at the atomic scale in the cryogenic regime. Since the acoustic vibration modes can be excited with the help of piezoelectric transducers or using Raman or Brillouin light scattering, the radiative heat exchanges could, in principle, be actively tuned at the atomic scale.
Future developments in this direction will certainly benefit  from recent progress in instrumental opto-mechanics~\cite{Clivia}.

\begin{acknowledgements}
This research was supported by the French Agence Nationale de la Recherche (ANR), under grant ANR-20-CE05-0021-01 (NearHeat). 
\end{acknowledgements}

\end{document}


\title{Supplemental material: Radiative heat exchange driven by the acoustic vibration\\modes between two solids at the atomic scale}

\author{M. G\'omez~Viloria}
\affiliation{Laboratoire Charles Fabry, UMR 8501, Institut d'Optique, CNRS, Universit\'{e} Paris-Saclay, 2 Avenue Augustin Fresnel, 91127 Palaiseau Cedex, France.}

\author{Y. Guo}
\affiliation{Institut Lumière Matière, Universit\'e Claude Bernard Lyon 1, CNRS, Universit\'e de Lyon, 69622 Villeurbanne, France.}

\author{S. Merabia}
\affiliation{Institut Lumière Matière, Universit\'e Claude Bernard Lyon 1, CNRS, Universit\'e de Lyon, 69622 Villeurbanne, France.}

\author{R. Messina}
\affiliation{Laboratoire Charles Fabry, UMR 8501, Institut d'Optique, CNRS, Universit\'{e} Paris-Saclay, 2 Avenue Augustin Fresnel, 91127 Palaiseau Cedex, France.}

\author{P. Ben-Abdallah}
\affiliation{Laboratoire Charles Fabry, UMR 8501, Institut d'Optique, CNRS, Universit\'{e} Paris-Saclay, 2 Avenue Augustin Fresnel, 91127 Palaiseau Cedex, France.}


\maketitle

We describe (1) the main steps to calculate the dielectric permittivity of a polar material by molecular dynamic simulation,  (2)  the physical origin for the contribution of acoustic modes in the optical response of materials in the short wavelengths limit and (3) the power law scaling of the thermal conductance between two MgO parallel plates with respect to their separation distance.

\section{Atomistic simulation of the dielectric response in polar materials}

We calculate the wave-vector and frequency-dependent nonlocal dielectric function of MgO by equilibrium molecular dynamics (EMD) based on the fluctuation-dissipation theorem, i.e. Eq. (4) in the main text. The molecular dynamics simulation is implemented in the LAMMPS package \cite{thompson} with the following empirical potential~\cite{matsui}:
\begin{equation}
\label{eq:phi}
	\phi_{ij}=\frac{q_iq_j}{r_{ij}}-\frac{C_iC_j}{r_{ij}^6}+f(B_i+B_j)\exp\left(\frac{A_i+A_j-r_{ij}}{B_i+B_j}\right),
\end{equation}
where $q_i$  and $q_j$   are the partial charges of atom $i$ and atom $j$ respectively, and  $r_{ij}$ is the distance between atom $i$ and atom $j$, $f$ is a standard force of 4.184 kJ/\AA /mol, the empirical energy parameters $A_i,B_i,C_i$   of atom $i$ can be found in the Ref.~\cite{matsui}. This empirical potential is chosen since it well reproduces the experimental data of frequency-dependent local dielectric function of MgO in the infrared regime~\cite{chalopin}. The first term in Eq.~\eqref{eq:phi} denotes the long-range Coulomb interaction between charged ions, whereas the second and third terms are the short-range bonding interaction. A cut-off distance of 10~\AA{} is adopted for the short-range interaction, and the particle-particle particle-mesh (pppm) method is used for treating the long-range Coulomb interaction.

A $10 \times 10 \times 10$ supercell of 8000 atoms is used in EMD simulation with a time step of 0.5 fs. Firstly the system is equilibrated under a $NVT$ (canonical) ensemble for 1 million steps. Then the system is switched to $NVE$ (microcanonical) ensemble for 5 million steps, where the system polarization density is sampled every 20 steps as calculated below:
\begin{equation}
	\mathbf P(t)=\frac1V\sum_i q_i \mathbf u_i(t)
\end{equation}
with $\mathbf u_i=\mathbf r_i -\mathbf r_{i,0}$  the displacement of atom $i$, and $V$ the total volume of the system. To reduce the statistical uncertainty, 10 independent EMD simulations are run for averaging. Once the time-dependent polarization density data is obtained, the Laplace-Fourier transform in Eq. (4) in the main text is calculated with a correlation length of 500 ps. According to the definition of the polalization density, the correlation function of dipolar moments can be related the correlation function of the displacement $\mathbf u_i$ of atoms
and by applying the fluctuation-dissipation theorem~\cite{Callen} we finally get
\begin{equation}
   \langle u_{i}^* u_{j} \rangle = 2\hbar\biggl(n + \frac{1}{2}\biggr) S(\omega) \delta_{ij} 
  \label{Eq:FDT}
\end{equation}
with $n(\omega,T)={1}/[\exp(\hbar\omega/k_{\rm B} T)-1]$ the Bose-Einstein distribution function at temperature $T$
and $S(\omega)$ is the spectrum of the displacement correlations at equilibrium. 

\section{Origin of the dipolar response induced by the acoustic vibration modes}
\begin{figure*}
	\includegraphics[width=0.6\linewidth]{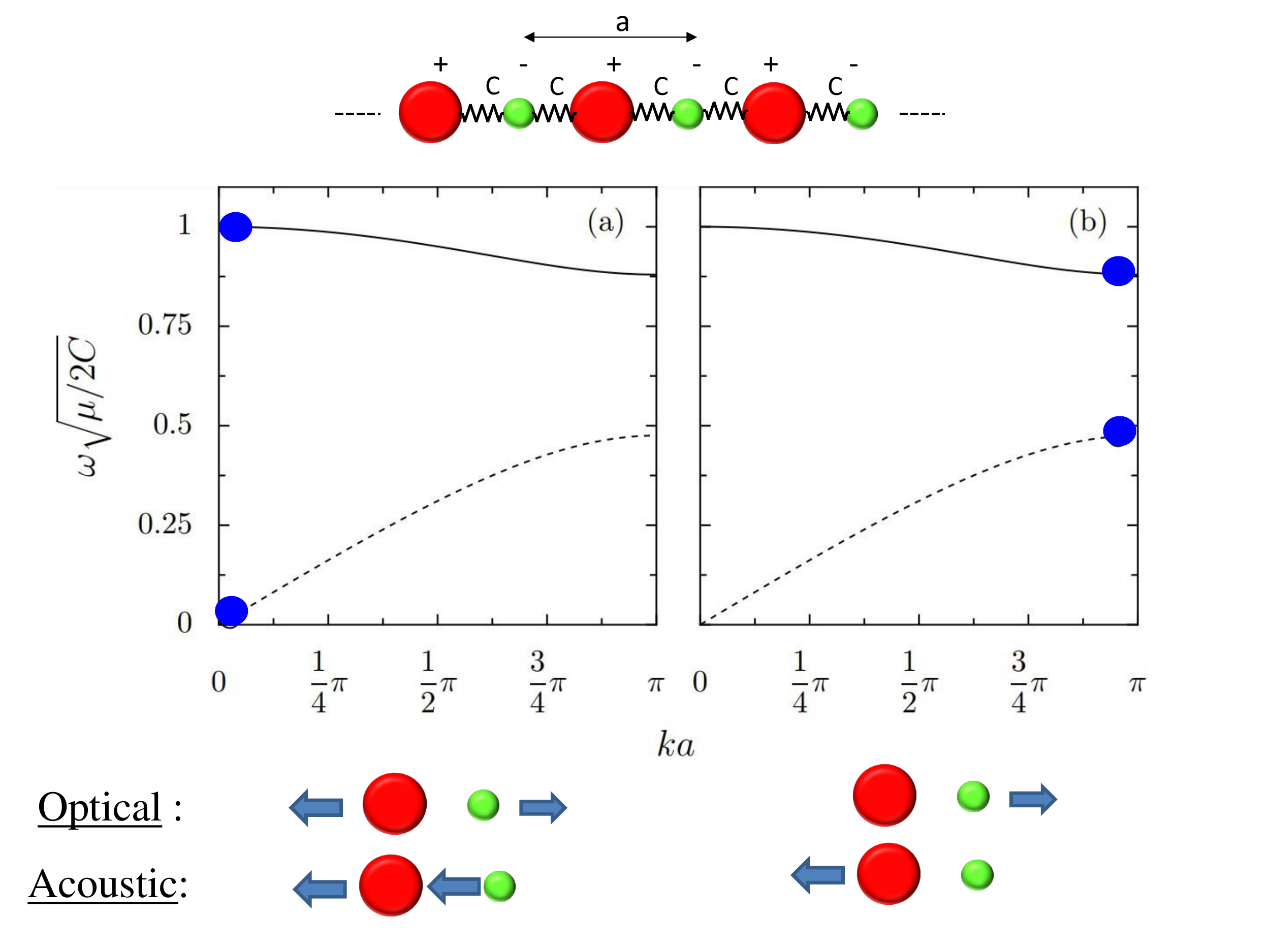}
	\caption{Optical behavior of a diatomic chain of period $a$ (top) at the edges of the Brillouin zone. Dispersion relation of eigenmodes (a) in the long wavelengths limit  (i.e. $k\sim 0$) and (b) in the short wavelengths limit (i.e. $k\sim \frac{\pi}{a}$). On the bottom the motion of heavy (red) and ligth (green) atoms both for the optical and acoustic vibration modes is shown. $C$ denotes the coupling strength between two neighbor atoms and $\mu$ is the reduced mass.}
	\label{Fig1}
\end{figure*}

In Ref.~[18] of the main text, the atomistic approach adopted by the authors is general and, in principle, it can be used to calculate precisely the heat flux exchanged between two polar materials. But the interpretation of physical results they give in their work is incomplete and even partially wrong. Indeed, the enhancement of heat flux they observed is attributed to the tunneling of acoustic phonons and not to the dipolar moments induced by these modes.

Here below we detail the physical origin of dipolar moments induced by the acoustic modes  in the particular case of a diatomic chain as sketched in Fig.~\ref{Fig1}. In the long wavelength limit [Fig.~\ref{Fig1}(a)] the motion of light and heavy atoms vibrate in phase opposition for the optical vibration modes  and they  vibrate in phase for acoustic  modes. Hence only the  optical modes give rise to dipole moments and are able to couple with an external electromagnetic field. 
On the contrary, in the short wavelength limit [Fig.~\ref{Fig1}(b)] both optical and the acoustic vibration modes give rise to dipole moment.  
In the case of  optical  modes the heavy atoms are fixed while the light atoms oscillate around their equilibrium position while for the acoustic modes the light atoms which are fixed and the heavy atoms oscillate as illustrated in the bottom of Fig.~\ref{Fig1}. 
\begin{figure*}
	\includegraphics[width=0.6\linewidth]{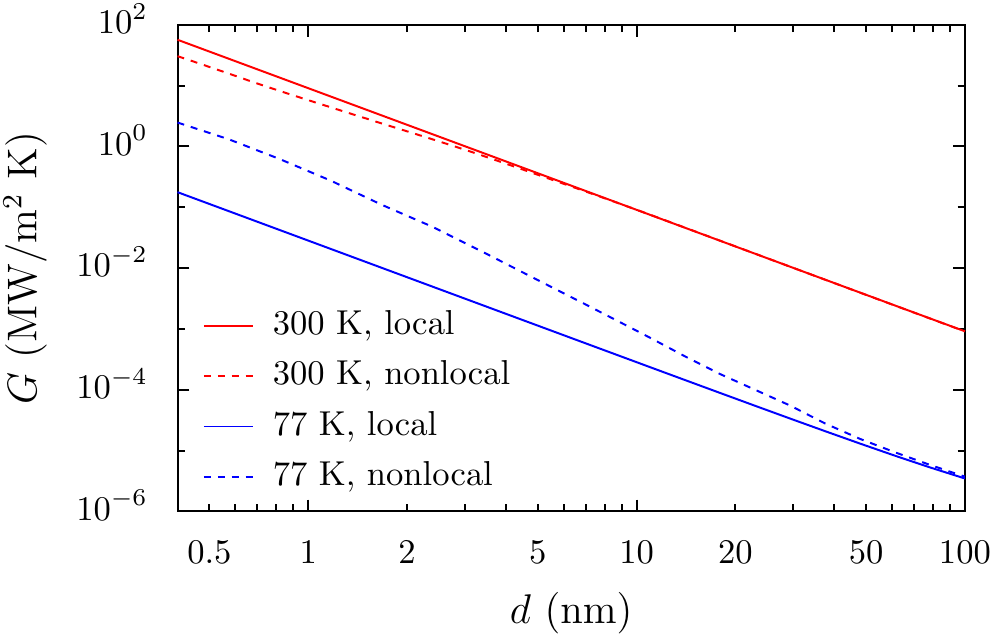}
	\caption{Heat-transfer coefficient at room temperature (300 K) as a function of the gap size for two infinite thick parallel plates made of MgO. The different lines correspond to the total contribution and to the contributions of propagating and evanescent waves for both TE and TM polarizations.}
	\label{Fig2}
\end{figure*}
This constitutes a radical change of paradigm from the state of the art where the acoustic vibrations modes are usually assumed to be purely mechanical modes and unable to couple to the electromagnetic field. Here we have demonstrated that this vision proves to be wrong in the short wavelength limit.  Beside this result, one of the main advantages of our approach in comparison with the one of Ref.~[18] is that we can quantitatively and qualitatively study the role played by the acoustic and optical modes in the heat transfer.

\section{Power law scaling of heat transfer vs the separation distance}
With polar materials the near-field heat transfer between two plates is dominated by TM evanescent waves that
can be shown to stem from surface phonon polaritons. 
These surface electromagnetic waves are hybrid or cavity modes that reside in both plates and have a penetration depth that is of the order of the gap size. For a given separation gap $d$ the cutoff wavevector is proportional to $1/d$~\cite{pba2010} leading to a $1/d^2$ dependence of the heat transfer coefficient. This power law scaling appears clearly on the simulations plotted in Fig.~\ref{Fig2}. However at cryogenic temperature the power law scaling changes since the transfer is not mediated anymore by the surface phonon-polariton.